\documentclass[manuscript]{emulateapj}

\def\Lsun{\hbox{L$_{\odot}$}}
\def\Msun{\hbox{M$_{\odot}$}}
\def\mmicron{\micron}
\def\mmicroJy{{\ensuremath{\mu}}Jy}
\newcommand{\degree}{\ensuremath{^\circ}}
\newcommand{\kms}{km~s$^{-1}$}

\begin{document}


\title{A Bright Submillimeter Source in the Bullet Cluster (1E0657-56) Field
Detected with BLAST}

\author{
        Marie~Rex,\altaffilmark{1,\dag}
        Peter~A.~R.~Ade,\altaffilmark{2}
        Itziar~Aretxaga,\altaffilmark{3}
        James~J.~Bock,\altaffilmark{4,5}
        Edward~L.~Chapin,\altaffilmark{6}
        Mark~J.~Devlin,\altaffilmark{1}
        Simon~R.~Dicker,\altaffilmark{1}
        Matthew~Griffin,\altaffilmark{2}
        Joshua~O.~Gundersen,\altaffilmark{7}
        Mark~Halpern,\altaffilmark{6}
        Peter~C.~Hargrave,\altaffilmark{2}
        David~H.~Hughes,\altaffilmark{3}
        Jeff~Klein,\altaffilmark{1}
        Gaelen~Marsden,\altaffilmark{6}
        Peter~G.~Martin,\altaffilmark{8,9}
        Philip~Mauskopf,\altaffilmark{2}
        Alfredo~Monta\~{n}a,\altaffilmark{3}
        Calvin~B.~Netterfield,\altaffilmark{9,10}
        Luca~Olmi,\altaffilmark{11,12}
        Enzo~Pascale,\altaffilmark{2}
        Guillaume~Patanchon,\altaffilmark{13}
        Douglas~Scott,\altaffilmark{6}
        Christopher~Semisch,\altaffilmark{1}
        Nicholas~Thomas, \altaffilmark{7}
        Matthew~D.~P.~Truch,\altaffilmark{1}
        Carole~Tucker,\altaffilmark{2}
        Gregory~S.~Tucker,\altaffilmark{14}
        Marco~P.~Viero,\altaffilmark{9}
        Donald~V.~Wiebe\altaffilmark{6,10}
}

\altaffiltext{1}{Department of Physics \& Astronomy, University of
Pennsylvania, 209 South 33rd Street, Philadelphia, PA 19104}
\altaffiltext{2}{Department of Physics \& Astronomy, Cardiff University, 5
The Parade, Cardiff, CF24~3AA, UK}
\altaffiltext{3}{Instituto Nacional de Astrof\'isica \'Optica y
Electr\'onica (INAOE), Aptdo. Postal 51 y 72000 Puebla, Mexico}
\altaffiltext{4}{Jet Propulsion Laboratory, Pasadena, CA 91109-8099}
\altaffiltext{5}{Observational Cosmology, MS 59-33, California Institute of Technology, Pasadena, CA 91125}
\altaffiltext{6}{Department of Physics \& Astronomy, University of
British Columbia, 6224 Agricultural Road, Vancouver, BC V6T~1Z1,
Canada}
\altaffiltext{7}{Department of Physics, University of Miami, 1320 Campo Sano Drive, Carol Gables, FL 33146}
\altaffiltext{8}{Canadian Institute for Theoretical Astrophysics, University of Toronto, 60 St. George Street, Toronto, ON M5S~3H8, Canada}
\altaffiltext{9}{Department of Astronomy \& Astrophysics, University of Toronto, 50 St. George Street, Toronto, ON  M5S~3H4, Canada}
\altaffiltext{10}{Department of Physics, University of Toronto, 60 St. George Street, Toronto, ON M5S~1A7, Canada}
\altaffiltext{11}{Istituto di Radioastronomia, Largo E. Fermi 5, I-50125, Firenze, Italy}
\altaffiltext{12}{University of Puerto Rico, Rio Piedras Campus, Physics Dept., Box 23343, UPR station, San Juan, Puerto Rico}
\altaffiltext{13}{Laboratoire APC, 10, rue Alice Domon et L{\'e}onie Duquet 75205 Paris, France}
\altaffiltext{14}{Department of Physics, Brown University, 182 Hope Street, Providence, RI 02912}
\altaffiltext{\dag}{\url{madamson@physics.upenn.edu}}

\slugcomment{To appear in the Astrophysical Journal}

\begin{abstract}
We present the 250, 350, and 500\,\mmicron\ detection of bright submillimeter
emission in the direction of the Bullet Cluster measured by the Balloon-borne
Large-Aperture Submillimeter Telescope (BLAST).  The 500 \mmicron\ centroid is
coincident with an AzTEC 1.1\,mm point-source detection at a position close to the 
peak lensing magnification produced by the cluster.  However, the 250\,\mmicron\ and
350\,\mmicron\ centroids are elongated and shifted toward the south with a
differential shift between bands that cannot be explained by pointing
uncertainties.  We therefore conclude that the BLAST detection is likely
contaminated by emission from foreground galaxies associated with the Bullet
Cluster. The submillimeter redshift estimate based on 250--1100\,\mmicron\
photometry at the position of the AzTEC source is $z_{\rm
phot}=2.9^{+0.6}_{-0.3}$, consistent with the infrared color redshift estimation
of the most likely IRAC counterpart.  These flux densities indicate an apparent
far-infrared luminosity of $L_{\rm FIR} =2 \times 10^{13}$\,\Lsun.  When the
amplification due to the gravitational lensing of the cluster is removed, the
intrinsic far-infrared luminosity of the source is found to be $L_{\rm FIR} \le
10^{12}$\,\Lsun, consistent with typical luminous infrared galaxies.
   
\end{abstract}


\section{Introduction}     
\label{sec:intro}

A decade of advances in submillimeter surveys has led to a broad understanding of the galaxies
responsible for the high-redshift component of the cosmic infrared background
\citep{Smail1997,Barger1998,Hughes1998,Webb2003,Greve2004,Pope2006,Coppin2006,Bertoldi2007,Knudsen2008,Devlin2009}.
These massive submillimeter galaxies (SMGs) are considered to be the products of galaxy
mergers and are associated with a vital stage of galaxy evolution.  With typical far-infrared
(FIR) luminosities $> 10^{12}$ \Lsun, they are presumed to be the high-redshift counterparts
to (ultra)luminous infrared galaxies (LIRGs, ULIRGs). The source of this energy is generally
attributed to powerful starbursts, suggesting star formation rates of 100--1000
\Msun\,yr$^{-1}$.  Approximately half of these galaxies are located at $1.9 \lesssim z
\lesssim 2.9$ \citep{Chapman2005,Aretxaga2007}, constituting a significant component of the 
star formation rate at this epoch \citep{Perez-Gonzalez2005}.  

The increased mapping speeds of current submillimeter instruments have
facilitated recent large-area blank-field surveys \citep{
Siringo2009,Devlin2009,Austermann2009}.  These surveys provide a large statistical sample of
bright SMGs.  However, resolution restricts the depth to the confusion limit and
they are unable to probe the abundant faint end of the number counts.
Observations directed toward massive galaxy clusters take advantage of
gravitational lensing to penetrate the confusion limit and sample intrinsically
fainter and higher-redshift SMGs \citep{Smail1997, Zemcov2007,Knudsen2008}.  

An extremely bright millimeter source was recently reported from a survey of
the Bullet Cluster conducted with the 1.1\,mm AzTEC receiver on the Atacama
Submillimeter Telescope Experiment (ASTE) \citep{Wilson2008}.  Its millimeter
properties are given in Table~\ref{table:aztec}. The high apparent flux and
close proximity to a critical line of magnification on the lensing model
motivated the authors to conclude that the source is likely to be a background
galaxy amplified by the foreground mass.  This AzTEC source
(MMJ065837$-$5557.0) is coincident with a LIRG detected by IRAC (which possibly
has multiple infrared images of the galaxy), with a redshift $z_{\rm phot}^{\rm
IR}=2.7\pm 0.2$ \citep{Wilson2008,Gonzalez2009}.

We present the 250, 350, and 500\,\mmicron\ measurements of a bright source
(BLAST~J065837$-$555708) located in the direction of the Bullet Cluster.  With significances
6-$\sigma$, 5-$\sigma$, and 9-$\sigma$ at 250, 350, and 500\,\mmicron, respectively, this
detection is the highest signal-to-noise source in the 0.25 deg$^{2}$ map of the region taken
with the Balloon-borne Large-Aperture Submillimeter Telescope (BLAST) in 2006
\citep{Pascale2008}. Details of the observations and data reduction techniques are presented
in Section~\ref{sec:obs}.  Extracted photometry as well as a photometric redshift estimate and
constraints on the intrinsic properties of the millimeter source are given in
Section~\ref{sec:anal}.  Finally, our conclusions are summarized in Section~\ref{sec:concl}.

\section{Observations and Reduction}
\label{sec:obs}
BLAST conducted an 11 day high-altitude flight from McMurdo Station, Antarctica,
in December of 2006 (BLAST06).  The telescope has a 1.8\,m primary mirror and
three large-format bolometer arrays that image the sky simultaneously at 250,
350, and 500\,\mmicron\ over a $14\arcmin \times 7\arcmin$ field of view.  The
telescope point spread function (PSF) at each wavelength is best fit by Gaussians
with full-widths at half-maximum (FWHM) 36\arcsec, 42\arcsec, and 60\arcsec\ at
250, 350, and 500\,\mmicron, respectively. BLAST scans the sky quickly ($\sim$
0.1\,deg s$^{-1}$) in azimuth, while drifting slowly in elevation.  Details of
the BLAST instrument can be found in \citet{Pascale2008}. The BLAST wavelengths
were selected to sample the peak of the spectral energy distribution for
high-redshift starburst galaxies as well as cooler star-forming regions within
our own Galaxy.

A large portion of the BLAST06 flight ($>$ 150/220 hours) was dedicated to
conducting large-area, wide and deep blank-field surveys to identify and
characterize the high-redshift population of starburst galaxies that comprise the
FIR extra-galactic background \citep{Devlin2009}. Two smaller fields were also
targeted which are known to contain massive galaxy clusters, including a
1.1\,deg$^2$ map around Abell 3112 and a 0.25\,deg$^2$ map around the Bullet
Cluster.  The Bullet Cluster field was visited three times during the flight,
with a total observing time of 2.0 hours.


The raw BLAST06 time-stream data are reduced using a common pipeline detailed in
\citet{Pascale2008} and \citet{Truch2008}.  The absolute calibration of the
BLAST06 data is determined based on observations of the star VY CMa with
10\%--13\% uncertainty which is highly correlated between bands
\citep{Truch2009}.  Signal and variance maps of the Bullet Cluster with
10\arcsec\ pixels are generated using the mapmaking algorithm OptBin
(Pascale et al., in prep).  The rotation of the sky between visits to the field
provided $\sim10\degree$ cross-linking between azimuthal scans.  This shallow
angle is insufficient to constrain low frequency drifts in the detectors and
preliminary maps suffered from significant striping along the telescope scan
direction. In order to suppress these artifacts, the time-stream data are
filtered by the map maker, whitening the map on scales $>$ 16\arcmin.  The
appropriate PSF for each band is produced by stacking maps of VY Cma, co-added
with the the proper sky rotation, and filtered in the same manner as the final
maps to enable accurate source extraction and calibration\footnote[1]{Maps and
PSFs for the field are available to download from http://blastexperiment.info}.
A catalog of sources for each band is compiled using a source-finding algorithm
which selects the peaks in a smoothed map produced by the noise-weighted
convolution of the image with the PSF.  The peak positions in each band
associated with BLAST~J065837$-$555708 are given in Table~\ref{table:pos}.

The telescope pointing solution is reconstructed post-flight by
combining 100\,Hz rate gyroscope data with $\sim$ 1\,Hz star camera
solutions which provide an absolute reference \citep{Pascale2008}.
The star cameras are mounted to the same rigid structure as the
telescope.  Nevertheless, tilting the structure in elevation results
in small changes to the angular offset between the boresights.  To
correct this effect, bright compact objects were targeted as pointing
calibrators at different elevations throughout the flight.  The
resulting 1-$\sigma$ pointing uncertainty over the elevation range of
observed calibrators (30\degree--50\degree) is $\sim$ 5\arcsec.
Observations of the Bullet Cluster were performed at high elevation
($\sim 54\degree$), outside the range of our pointing calibrators.
They were also conducted early in the flight, before the telescope
focus had been finalized. While these factors lead us to suspect
larger absolute pointing uncertainties for sources in the Bullet
Cluster field, the optical design of BLAST does not allow for relative
offsets between the bands.

The absolute pointing offset for the BLAST maps of the Bullet Cluster is
determined by stacking them against positions of 24\,\mmicron\ sources.  This
24\,\mmicron\ catalog contains sources with a significance $\ge$ 3-$\sigma$
extracted from the {\it Spitzer} MIPS observations of the field using the APEX
source-finding algorithm \citep{Makovoz2002}.  At 250\,\mmicron, this analysis
produces the highest signal-to-noise, indicating a 27\arcsec\ astrometry
correction required for the BLAST maps.  The corrected peak positions of 
BLAST~J065837$-$555708 are given in Table~\ref{table:pos}.

\section{Analysis} 
\label{sec:anal}

\begin{figure*}
\centering
\includegraphics[width=\linewidth]{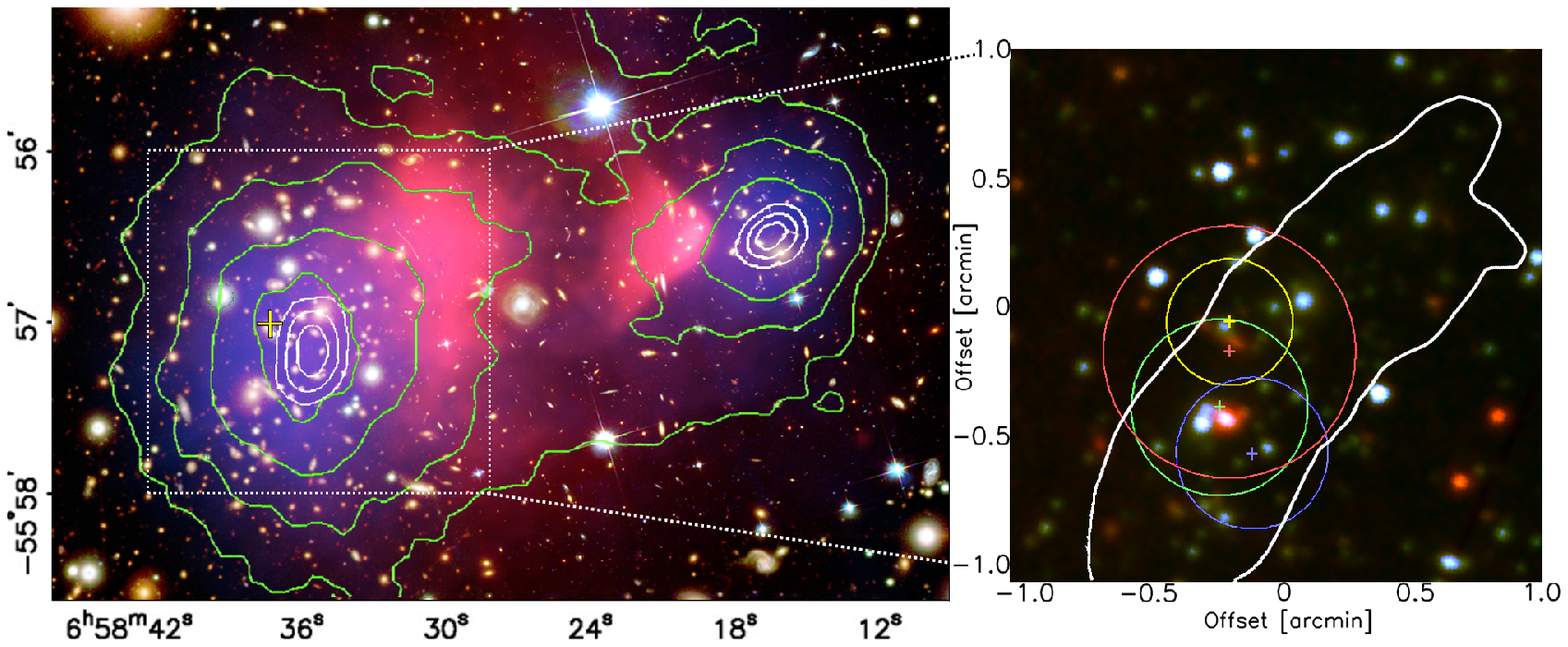}
\caption{{\it Left panel:} The background image combines optical maps from
Magellan and {\it HST} \citep{Clowe2004} with {\it Chandra} X-ray data
\citep{Markevitch2002}. The X-ray map reveals the hot gas which represents the
majority of the baryonic matter in the cluster (shown in pink). The distribution
of the dark matter is shown in blue, with weak lensing contours over-plotted in
green. Image based on Figure~1 in \citet{Clowe2006}.  The yellow cross indicates
the position of the AzTEC detection.  
{\it Right panel:} A 2\arcmin $\times$ 2\arcmin\ detail based on IRAC 3.6, 4.5,
and 8.0~\mmicron\ maps of the region.  The BLAST 250, 350, and 500\,\mmicron\
positions are shown in blue, green, and red, respectively.  The AzTEC 1.1\,mm
position is shown in yellow.  The crosses indicate the centroids of the
detections and the circles represent the FWHM of each beam. The white contour
denotes the critical curve corresponding to infinite magnification based on {\it
HST} data \citep{Bradac2006}.   
}
\label{fig:maps} 
\end{figure*}

The Bullet Cluster, at redshift $z = 0.297$ \citep{Tucker1998}, is the product
of two merging galaxy clusters.  The less massive member has passed through the
dominant mass with a relative velocity of $\gtrsim$ 4500\,\kms.  This extreme
interaction emphasizes the displacement between the mass of the cluster, traced
by weak lensing models,  and the hot X-ray gas, providing dramatic evidence for
the presence of dark matter \citep{Clowe2004}. Figure~\ref{fig:maps} shows maps
of the Bullet Cluster at optical, X-ray, and infrared wavelengths.  The image in
the left-hand panel combines optical maps from Magellan and the {\it Hubble Space
Telescope} (HST) \citep{Clowe2004} with {\it Chandra} X-ray data
\citep{Markevitch2002}.  The X-ray emission reveals hot gas that represents the
majority of the baryonic matter in the cluster (shown in pink).  The distribution
of the dark matter is shown in blue with weak lensing contours over-plotted in
green \citep{Clowe2006}.  The yellow cross marks the position of the AzTEC point-source
detection at 1.1\,mm, near the peak of the largest mass distribution.  

The right-hand panel of Figure~\ref{fig:maps} shows a  $2\arcmin \times 2\arcmin$
detail of the region.  The color image is produced from {\it Spitzer} IRAC 3.6,
4.5, and 8.0\,\mmicron\ maps.  The BLAST centroids at 250, 350, and 500\,\mmicron\ 
are shown in blue, green, and red, respectively.  The AzTEC 1.1~mm position is
shown in yellow.  Each cross indicates the position of the peak BLAST detection, with
circles representing the FWHM of each beam.  It is apparent that the relative
offsets between the BLAST beams is inconsistent with the 5\arcsec\ pointing error
determined in other BLAST06 maps \cite{Pascale2008, Devlin2009}.  We therefore consider the
possibility that the centroids are shifted due to emission from a combination of
galaxies at different redshifts, slightly displaced along the line-of-sight. 

\subsection{Photometry}
\label{sec:scen2}

\begin{figure}
\centering
\includegraphics[width=\linewidth]{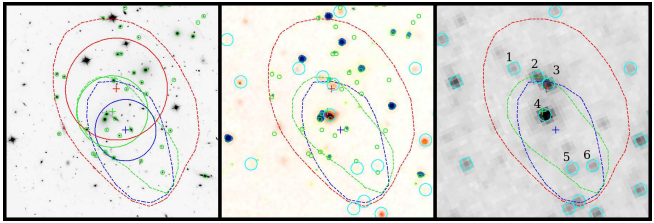}
\caption{Detailed 2\arcmin $\times$
2\arcmin\ region around BLAST~J065837$-$555708. The BLAST 250, 350, and
500\,\mmicron\ positions are shown in blue, green, and red, respectively.  
Crosses indicate the centroids of the measurements and corresponding solid circles 
represent the FWHM of the beam. Dashed lines show 3-$\sigma$ contours for
each BLAST detection. 
{\it Left-hand panel:} {\it HST} 606\,nm map.  Small green circles indicate the
positions of spectroscopically verified cluster members \citep{Mehlert2001}.
Many of these foreground galaxies may be contributing flux to the
submillimeter measurement and shifting the centroid positions away from the
position of the background LIRG. 
{\it Middle panel:} Red, green, and blue composite image made from IRAC 8.0, 4.5, and
3.6~\mmicron\ maps, respectively. Again, cluster members are indicated with small green
circles.  24\,\mmicron\ counterparts detected with a significance $>$ 5-$\sigma$ are marked in
cyan.  The infrared maps reveal a number of bright galaxies near the peak of the
350\,\mmicron\ emission.  {\it Right-hand panel:}  MIPS 24\,\mmicron\ map.  Cyan circles
indicate $>$ 5-$\sigma$ sources.  Sources within the 3-$\sigma$ BLAST contours are numbered
and the extracted infrared photometry is given in Table~\ref{table:photom}.  Sources 2 and 3
are likely split images of the lensed background LIRG associated with the 1.1\,mm detection.
They coincide with the peak of the 500\,\mmicron\ emission, although the shorter wavelength
peaks fall closer to source 4.  This lower redshift source is likely to be brighter at short
BLAST wavelengths, contributing significantly to the emission in those bands.  Sources 1, 5,
and 6, which also appear at 24\,\mmicron\ may additionally contribute at BLAST wavelengths.
The elongation of the contours could be explained by confusion of these sources in the lower
resolution BLAST maps.
}
\label{fig:thumbs}
\end{figure}

The displacement of the BLAST centroid positions from the location of the millimeter
source increases with frequency  and detections are elongated in the direction of that
displacement.  These features are illustrated in Figure~\ref{fig:thumbs}, showing a
detailed 2\arcmin $\times$ 2\arcmin\ region around BLAST~J065837$-$555708. The
differential shift between bands suggests emission from other galaxies with small
angular offsets to the line-of-sight, south of the AzTEC source.  The central panel of
Figure~\ref{fig:thumbs} is a composite of IRAC 3.6, 4.5, and 8.0\,\mmicron\ images.  The
maps reveal multiple cluster member galaxies that may be contributing flux to the BLAST
submillimeter measurements and shifting the centroid positions away from the background
LIRG.  For a galaxy contained within the Bullet Cluster ($z \sim 0.3$), BLAST colors
sample the Rayleigh-Jeans side of the SED.  This emission scales as $S_{\lambda} \sim
\lambda^{-4}$, and would contribute to the BLAST bands in a form consistent with the
observations.  The MIPS 24\,\mmicron\ map is shown in the right-hand panel of
Figure~\ref{fig:thumbs}, with cyan circles around $>$ 5-$\sigma$ sources.  Those located
within the 3-$\sigma$ flux density contours of the BLAST emission are numbered and
infrared photometry extracted for each using the APEX pipeline is given in
Table~\ref{table:photom}.  Though a number of these galaxies likely contribute to the
submillimeter detection, sources 2, 3, and 4 clearly dominate the emission at
24\,\mmicron\ and line up with the centroid peaks of the submillimeter maps.  Sources 2
and 3 are likely double images of the lensed background LIRG \citep{Wilson2008,Gonzalez2009},
while Source 4 is a spectroscopically confirmed cluster member \citep{Mehlert2001}.  
We estimate the submillimeter flux densities of each of the lensed high-redshift
millimeter source detected by AzTEC and the foreground dominant cluster member galaxy by
simultaneously fitting the PSF to the BLAST maps at the positions of both sources using
the maximum-likelihood algorithm presented in \citet{Scott2002}. Because the
mid-infrared images of the lensed LIRG are closely spaced compared to the scale of the 
resolution, we adopt the millimeter position of Wilson et al (2008) for this component in
the joint-fit. The submillimeter photometry of the LIRG and the bright cluster member
galaxy are given in Table~\ref{table:data} and Table~\ref{table:data2} respectively.
The measured 500 \mmicron\ emission is dominated by an association with the millimeter AzTEC
source. However, our estimates of the shorter wavelength contributions to the flux
densities at 250 and 350 \mmicron\ favor the cluster member galaxy as the dominant source
of  emission.  Consequently, the observed submillimeter spectrum of the background LIRG
is flattened across the BLAST bands.  
 
Submillimeter emission in this region is not confined to compact sources.  {\it
IRAS} maps reveal foreground cirrus emission of 5\,MJy\,sr$^{-1}$ at
100\,\mmicron. However, this structure is smooth on scales of $\lesssim 1\degree$
and therefore filtered out of the BLAST maps.  
The measured BLAST flux densities are adjusted to account for a small
contamination by the thermal Sunyaev-Zel'dovich effect (SZE). This excess flux
density results from the hot gas in the cluster boosting cosmic microwave
background photons to higher energies through inverse Compton scattering.
Following \citet{Wilson2008}, we estimate the SZE flux contribution to the BLAST
bands at the position of the AzTEC detection by developing a model to describe
the electron density distribution of the Bullet Cluster.  We assume a
two-component spherically-symmetric $\beta$-model ($\beta = 0.7$) and a range of
total cluster-mass, $M_{\mathrm {tot}} = 1.5$--$2.1 \times 10^{15}$\,\Msun,
distributed in a 3:1 ratio \citep{Nusser2008} between the main and sub-clusters.
Each component is centered at the peak of the corresponding observed X-ray
surface brightness distribution given by \citet{Clowe2006}.  Provided that the
electron temperature is similar to the X-ray gas temperature, and assuming each
component is isothermal, we estimate electron temperatures $\sim
15.4$--$16.3$\,keV  for the main cluster and $\sim 6.0$--$7.4$\,keV for the
sub-cluster, in agreement with the temperature map provided by
\citet{Markevitch2002}. Relativistic corrections were applied using the
analytical approximation given by \citet{Itoh1998}. The results of these
simulations and our revised source flux densities are given in
Table~\ref{table:data}.  We have also used this model to estimate the SZE at the 
position of the bright cluster member galaxy and these results are shown in 
Table~\ref{table:data2}.

\subsubsection{Photometric Redshifts}
\label{sec:photoz}

The shape of the spectral energy distribution (SED) in the submillimeter regime
can be used to estimate the redshift of the source. The BLAST bands were chosen
to optimize constraints on the FIR peak of the bright submillimeter galaxy
population at $1<z<4$.  A Monte Carlo photometric redshift method has been
developed \citep{Hughes2002, Aretxaga2003} to take into account constraining
prior information such as a range of local SED templates and the evolving
luminosity function to $z\approx 2$.

In accordance with this method, we generate SEDs for a catalog of galaxies based
on an evolutionary model for the 60\,\mmicron\ luminosity function that fits the
observed 850\,\mmicron\ number counts (e.g.~luminosity evolution $\propto
(1+z)^3$ for $z\lesssim 2$, and no evolution at $z>2$) and covers a simulated area
of $10\,{\rm deg}^{2}$.  Template SEDs are drawn at random from a library of
local starbursts, ULIRGs, and AGN to provide FIR--radio fluxes. The SEDs cover a
wide range of FIR luminosities ($10^{9.0}$--$10^{12.3}$\,\Lsun) and temperatures
(25--65\,K).  The photometric and calibration uncertainties of the BLAST and
AzTEC measurements are included in the fluxes of the mock galaxies.  We adopt a
conservative estimate for the amplification of the source ($A = 30$)
\citep{Wilson2008, Gonzalez2009}, and reject mock galaxies with fluxes below
the detection threshold of the survey.  The redshift probability distribution of
the source is then calculated as the normalized distribution of the redshifts of
the mock galaxies in the catalog, weighted by the likelihood of identifying the
colors and fluxes of each mock galaxy with those of the source.

\begin{center}
\begin{figure}
\includegraphics[width=\linewidth]{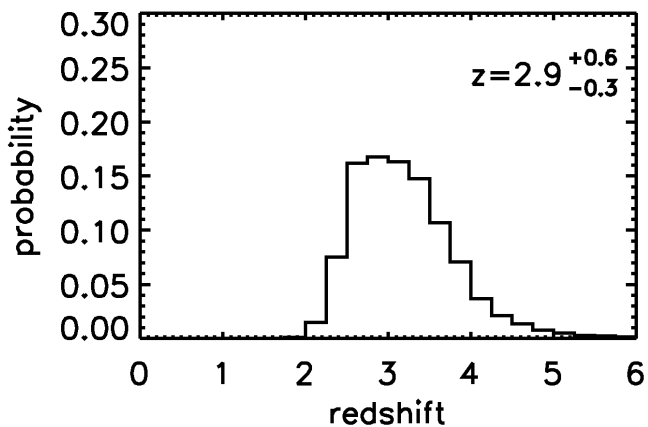}
\caption{The redshift probability distribution based on 250--1100\,\mmicron\
photometry at the millimeter centroid position.  The uncertainty in lensing
magnification results in a longer high-redshift tail.}
\label{fig:zphot2}
\end{figure}
\end{center}

Figure~\ref{fig:zphot2} shows the results of this analysis indicating a photometric
redshift $z_{\rm phot}= 2.9^{+0.6}_{-0.3}$.  This estimate is consistent with the
IR-color redshift estimation ($z_{\rm phot}^{\rm IR}=2.7\pm 0.2$) for the most likely
IRAC counterpart \citep{Wilson2008, Gonzalez2009}. The accuracy of the technique has
been experimentally tested for galaxies detected in at least 3 bands in the rest-frame
radio to FIR interval, yielding a typical 1-$\sigma$ error $\langle ( z^{\rm MC}_{\rm
phot}-z_{\rm spec})^2 \rangle ^{1/2} \approx 0.25$ over the $0.5\lesssim z \lesssim4.0$
regime \citep{Aretxaga2007}.

\subsubsection{Intrinsic Properties}
\label{sec:props}

\begin{figure}
\centering
\includegraphics[width=0.9\linewidth]{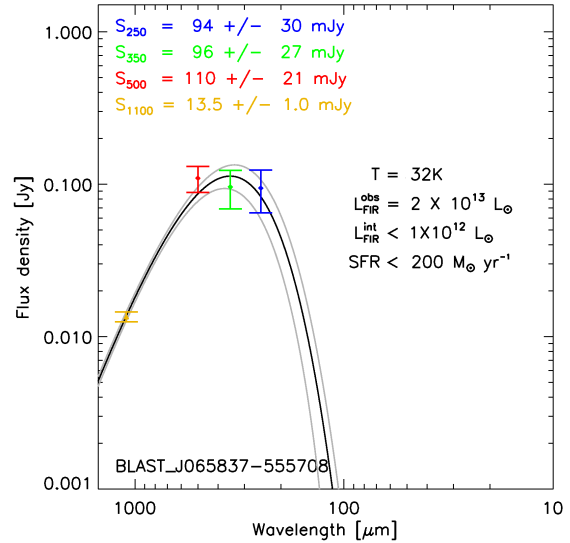}
\caption{Spectral Energy Distribution of BLAST~J065837$-$555708. The black curve
is the best-fit model to the BLAST and AzTEC measurements, shown on the plot with
1-$\sigma$ error bars.  These flux densities have been corrected for the
contribution from SZE. The grey curves indicate the 1-$\sigma$ range of
temperature and amplitudes for the fit.  The spectral index is fixed at $\beta =
1.5$.  The spectral information enables a color correction to the measured flux 
densities which accounts for the profiles of the BLAST filters. These color
corrected values are displayed on the plot.  $L_{\rm FIR}^{\rm obs}$ represents the 
far-infrared luminosity based on the fit to observations, while 
$L_{\rm FIR}^{\rm int}$ is an estimate of the intrinsic far-infrared luminosity of 
the source, assuming an amplification of $A > 30$ due to gravitational lensing.  
An upper limit on the star formation rate is derived based on this luminosity 
\citep{Kennicutt1998}.
}
\label{fig:sed2}
\end{figure}

With an estimate of its redshift, we can derive properties
intrinsic to the submillimeter galaxy.  Figure~\ref{fig:sed2} shows a
single-temperature modified blackbody SED of the form
\begin{equation}
\label{eqn:sed}
S_{\nu} = N(\nu/\nu_{0})^{\beta}B_{\nu}(T),
\end{equation}
where $S_{\nu}$ is the flux density, $N$ is the amplitude, $\beta$ is the dust
emissivity index, $\nu_{0}$ is fixed at $c/250$\,\mmicron, and $B_{\nu}(T)$ is the
Planck blackbody radiation function for a source with temperature $T$.  The function is
fit to the data using a $\chi^2$ minimization routine, after subtracting the
contribution from SZE and an estimate of the contribution from the brightest mid-infrared
cluster member galaxy, as discussed previously. The grey curves indicate the 1-$\sigma$
range of temperature and amplitudes for the fit estimated from 200 Monte Carlo
simulations.  The value of $\beta$ is fixed here at $\beta = 1.5$, but the fits give
very similar values of $T$ and $L_{{\rm FIR}}$ if we use $\beta = 2.0$. The best-fit
temperature of the distant lensed source is 32 K.  This SED is consistent with those fit to
submillimeter galaxies observed in regular blank-field surveys
\citep{Chapman2005,Kovacs2006,Aretxaga2007,Dye2009}. This spectral information also
enables a color correction to the flux densities which accounts for the profiles and
widths of the BLAST filters \citep{Truch2008}.  These color corrected flux densities are
given in Table~\ref{table:data}.  $L_{\rm FIR}^{\rm obs}$ is the FIR luminosity,
calculated in the wavelength range $\lambda= 8$--$1000$\,\mmicron\
\citep{Kennicutt1998}, for the observed flux densities without considering gravitational
lensing effects.  Using the previously assumed lower limit for the amplitude of the
lensing magnification ($A > 30$), we derive an upper limit for the intrinsic FIR
luminosity of $L_{\rm FIR}^{\rm int} < 10^{12}$ \Lsun.  This value is consistent with
estimates presented by the AzTEC team \citep{Wilson2008} as well as from extrapolation
of the 24\,\mmicron\ luminosity \citep{Gonzalez2009}, and qualifies the source as a
LIRG.  This intrinsic luminosity implies an upper limit on the star formation rate of
200 \Msun\,yr$^{-1}$ \citep{Kennicutt1998}.  The source is relatively faint compared to
typical submillimeter-selected galaxies with similar dust temperatures
\citep{Blain2003,Kovacs2006,Dye2009}, and is more consistent with the
luminosity-temperature relationship observed in IRAS-selected samples
\citep{Soifer1991,Blain2003,Chapman2003,Chapin2009}. 
\section {Conclusions}
\label{sec:concl}

BLAST submillimeter maps of the Bullet Cluster have revealed bright emission near
the peak of the weak lensing contours \citep{Clowe2006}.  The position of the
detection coincides with the AzTEC 1.1\,mm source MMJ065837$-$5557.0 and a
multiply-imaged LIRG detected in the {\it Spitzer} IRAC maps
\citep{Wilson2008,Gonzalez2009}.  The systematic offsets in peak emission between
the BLAST bands motivates the likely interpretation that multiple sources are
confused at BLAST resolution. We have performed a simultaneous fit to the
brightest two infrared galaxies in the field in order to estimate the properties
of the background LIRG.  The resulting color corrected flux densities for the the
source are $S_{250} = 94 \pm 30$, $S_{350} = 96 \pm 27$, and $S_{500} = 110 \pm
21$\,mJy.

The BLAST colors sample the peak of the source spectrum, and 250--1100\,\mmicron\ flux
densities place strong constraints on its photometric redshift.  Using a Monte Carlo
photometric redshift method \citep{Hughes2002, Aretxaga2003}, we have calculated a redshift
$z_{\rm phot}= 2.9^{+0.6}_{-0.3}$. The redshift probability distribution is inconsistent with
the redshift of the Bullet Cluster ($z \approx 0.3$), ruling out the possibility that the
millimeter AzTEC source is associated with the foreground mass.  Assuming the best-fit
photometric redshift we have derived the observed FIR luminosity of the source, in the absence
of lensing; to be $L_{\rm FIR}^{\rm obs} = 2 \times 10^{13}$\,\Lsun.  The proximity of the
source to a critical line in the lensing model \citep{Bradac2006,Gonzalez2009} puts a lower
limit $A > 30$ on the amplitude of magnification due to gravitational lensing by the cluster.
This imposes an upper limit on the intrinsic luminosity of the SMG, $L_{\rm FIR}^{\rm int} <
10^{12}$\,\Lsun.  This luminosity is consistent with estimates from \citet{Wilson2008} and
\citet{Gonzalez2009}, and characterizes the source as a LIRG, a member of the fainter
population of SMGs that dominate the number counts but are difficult to detect with the
limited resolution of existing submillimeter telescopes.   

The challenges in interpreting the BLAST maps demonstrate the necessity of including
complementary, higher resolution data sets in the analysis.  BLAST is a precursor to the
Spectral and Photometric Imaging Receiver (SPIRE) on the recently launched {\it Herschel
Space Observatory} \citep{Griffin2006}.  With nearly twice the resolution, and gains in
sensitivity, SPIRE will be able to provide significantly better centroiding accuracy.  
Nevertheless, access to infrared and millimeter data sets will still be integral to a more 
complete understanding of these crowded systems.

We acknowledge the support of NASA through grant numbers NAG5-12785, NAG5- 13301, and
NNGO-6GI11G, the NSF Office of Polar Programs, the Canadian Space Agency, the Natural
Sciences and Engineering Research Council (NSERC) of Canada, and the UK Science and
Technology Facilities Council (STFC).  We thank Tony Mroczkowski for his help with the
SZE simulations.  This research has been enabled by the use of WestGrid computing
resources. This research also made use of the SIMBAD database, observations made with the
Spitzer Space Telescope, which is operated by the Jet Propulsion Laboratory, California
Institute of Technology under a contract with NASA, and SAOImage DS9, developed by
Smithsonian Astrophysical Observatory. 

\begin{deluxetable}{lc}
\tablewidth{0pt}
\tablecaption{Millimeter properties of AzTEC source MMJ065837$-$5557.0. \label{table:aztec}}
\startdata
\cline{1-2}
Flux Density          & 13.5 $\pm$ 1.0\,mJy \\ 
RA Centroid Position  & $6^{\mathrm h}58^{\mathrm m}37.\!^{\mathrm s}31$ (J2000) \\
Dec Centroid Position & $-55\degree 57\arcmin 1\farcs 5$ (J2000) \\
\enddata
\tablecomments{1.1\,mm flux density and centroid position of MMJ065837$-$5557.0
from \citet{Wilson2008}.}
\end{deluxetable}
\begin{deluxetable*}{lcccc}
\tablewidth{0pt}
\tablecaption{Peak Positions of BLAST~J065837$-$555708. \label{table:pos}}
\tablehead{
\colhead{} &
\multicolumn{2}{c}{Before Correction} & 
\multicolumn{2}{c}{After Correction}  \\
\colhead{$\lambda$} & \colhead{Centroid RA} & \colhead{Centroid Dec} &
\colhead{Centroid RA} & \colhead{Centroid Dec} 
\\
\colhead{[$\mmicron$]}   & \colhead{[h:m:s]} & \colhead{[d:m:s]} &
\colhead{[h:m:s]} & \colhead{[d:m:s]} 
\\
}
\startdata
250 & 06:58:35.3 & $-$55:57:08 & 06:58:36.7 & $-$55:57:33 \\ 
350 & 06:58:36.2 & $-$55:56:58 & 06:58:36.6 & $-$55:57:22 \\
500 & 06:58:36.0 & $-$55:56:44 & 06:58:37.3 & $-$55:57:09 \\
\enddata
\tablecomments{Positions of the peaks in each band associated with
BLAST~J065837$-$555708.  The positions are given before and after the correctional
offset applied based on stacking the 250\,\mmicron\ map against 24\,\mmicron\
sources.}
\end{deluxetable*}
\begin{deluxetable*}{lccccccc}
\tabletypesize{\footnotesize}
\tablewidth{0pt}
\tablecaption{Infrared photometry for 24\,\mmicron\ sources associated with
BLAST~J065837$-$555708. \label{table:photom}}
\tablehead{
\colhead{ID} & \colhead{24\,\mmicron\ RA} & \colhead{24\,\mmicron\ Dec} &
\colhead{S$_{\rm 3.6}$}  &
\colhead{S$_{\rm 4.5}$}  &
\colhead{S$_{\rm 5.8}$}  &
\colhead{S$_{\rm 8.0}$}  &
\colhead{S$_{\rm 24}$}\\
\colhead{}   & \colhead{[Hours]} & \colhead{[Degrees]} &
\colhead{[\mmicroJy]}   &
\colhead{[\mmicroJy]}   &
\colhead{[\mmicroJy]}   &
\colhead{[\mmicroJy]}   &
\colhead{[\mmicroJy]}   \\
}
\startdata
1 & 6:58:39.69 & $-$55:56:56 & 18.6 $\pm$  0.2 & 26.2 $\pm$   0.3     & 
70.6  $\pm$   1.1     & 40.0   $\pm$   1.9 &   252.7 $\pm$ 37.2 \\
2 & 6:58:37.99 & $-$55:57:01 & 19.1 $\pm$  0.2 & 25.7 $\pm$   0.3     & 
76.5  $\pm$   1.1     & 60.5   $\pm$   1.9 &   596.2 $\pm$ 34.9 \\
3 & 6:58:37.21 & $-$55:57:06 & 28.5 $\pm$  0.2 & 49.0 $\pm$   0.3     & 
69.2  $\pm$   1.1     & 86.9   $\pm$   2.0 &   840.4 $\pm$  36.4 \\
4 & 6:58:37.40 & $-$55:57:24 & 309.6 $\pm$ 0.5 & 269.5$\pm$    0.5    &  
190.4 $\pm$   1.3     & 528.3  $\pm$   2.1 &   1482.0$\pm$  34.8 \\
5 & 6:58:35.62 & $-$55:57:55 & 72.0 $\pm$  0.2 & 47.4 $\pm$   0.3     &  
60.2  $\pm$  1.1     & 25.6   $\pm$   1.9  &   306.4 $\pm$  37.4 \\
6 & 6:58:34.09 & $-$55:57:54 & 32.5 $\pm$  0.2 & 23.0 $\pm$   0.3     &  
41.9  $\pm$  1.1     & 18.4   $\pm  $ 1.9  &   285.0 $\pm$ 36.5 \\
\enddata
\tablecomments{IR photometry corresponding to the 24\,\mmicron\ sources located 
within the BLAST 3-$\sigma$ contours which have a significance of $>$ 5-$\sigma$.}
\end{deluxetable*}
\begin{deluxetable*}{lcccc}
\tablewidth{0pt}
\tablecaption{Submillimeter Flux Densities of MMJ065837$-$5557.0. \label{table:data}}
\tablehead{
\colhead{$\lambda$} & \colhead{Measured Flux} &
\colhead{SZE} & \colhead{SZE Corrected Flux} &
\colhead{Color Corrected Flux} \\
\colhead{[\mmicron]} &  \colhead{[mJy]} &
\colhead{[mJy]} & \colhead{[mJy]} & \colhead{[mJy]} \\
}
\startdata
250 & 91 $\pm$ 27 &  0.15 $\pm $0.03 &  91  $\pm $27  &  94 $\pm $30\\
350 & 95 $\pm$ 24 &   1.1 $\pm $0.2  &  94  $\pm $24  &  96 $\pm $27\\
500 &121 $\pm$ 16 &  10.6 $\pm $1.7  &  110 $\pm $16  &  110 $\pm $21\\
\enddata
\tablecomments{Flux densities in each band associated with BLAST~J065837$-$555708
at the position of the AzTEC detection. The contribution from SZE is calculated
at the position of the AzTEC point source, based on the theoretical model outlined in 
\citet{Wilson2008}. This small contamination is subtracted from the measured fluxes for
further analysis. The monochromatic flux densities based on the best-fit SED are given in 
the final column.}
\end{deluxetable*}
\begin{deluxetable*}{lcccc}
\tablewidth{0pt}
\tablecaption{Submillimeter Flux Densities of the dominant Cluster Member Galaxy. \label{table:data2}}
\tablehead{
\colhead{$\lambda$} & \colhead{Measured Flux} &
\colhead{SZE} & \colhead{SZE Corrected Flux} &
\colhead{Color Corrected Flux} \\
\colhead{[\mmicron]} &  \colhead{[mJy]} &
\colhead{[mJy]} & \colhead{[mJy]} & \colhead{[mJy]} \\
}
\startdata
250 & 129 $\pm$ 26 &  0.13 $\pm $0.03 &  129  $\pm $26  &  131 $\pm $30\\
350 & 104 $\pm$ 23 &  1.0  $\pm $0.2  &  103  $\pm $23  &  106 $\pm $27\\
500 &  49 $\pm$ 18 &  9.4 $\pm  $1.5  &   40  $\pm $18  &   38 $\pm $18\\
\enddata
\tablecomments{Submillimeter flux densities in each band associated with the dominant
24\,\mmicron\ cluster member galaxy at RA (J2000) = 6:58:37.40, Dec (J2000) = $-$55:57:24,
identified as Source~$4$ in Figure~\ref{fig:thumbs} and Table~\ref{table:photom}.  The
columns are analogous to those in Table~\ref{table:data}.} 
\end{deluxetable*}


\clearpage

\bibliographystyle{plainnat}
\bibliography{refs}

\end{document}